\documentclass[final,5p,times,twocolumn]{elsarticle}

\usepackage{lineno}
\usepackage[hidelinks]{hyperref}
\linenumbers
\usepackage{url}
\usepackage{graphicx}
\usepackage{amsmath}   
\usepackage{xcolor}

\usepackage{lineno}
\nolinenumbers

\begin{document}

\begin{frontmatter}

\title{High-Precision Alignment Techniques for Realizing an Ultracompact Electromagnetic Calorimeters Using Oriented high-Z Scintillator Crystals}
\author[infnferrara]{Lorenzo Malagutti\corref{cor1}}
\cortext[cor1]{First author}
\author[como,bicocca]{Alessia Selmi\corref{mycorrespondingauthor}}
\cortext[mycorrespondingauthor]{Corresponding author}
\ead{aselmi@uninsubria.it}
\author[infnferrara]{Laura Bandiera}
\author[minsk]{Vladimir Baryshevsky}
\author[como,bicocca]{Luca Bomben}
\author[infnferrara]{Nicola Canale}
\author[como,bicocca]{Stefano Carsi}
\author[como]{Mattia Ciliberti}
\author[padova,legnaro]{Davide De Salvador}
\author[infnferrara,ferrara]{Vincenzo Guidi}
\author[minsk]{ Viktar Haurylavets}
\author[minsk]{Mikhail Korjik}
\author[como,bicocca]{Giulia Lezzani}
\author[minsk]{Alexander Lobko}
\author[como,bicocca]{Sofia Mangiacavalli}
\author[infnferrara]{Andrea Mazzolari}
\author[infntrieste,trieste]{Pietro Monti-Guarnieri}
\author[frascati]{Matthew Moulson}
\author[infnferrara,ferrara]{Riccardo Negrello}
\author[infnferrara]{Gianfranco Paternò}
\author[como,bicocca]{Leonardo Perna}
\author[como,bicocca]{Christian Petroselli}
\author[como,bicocca]{Michela Prest}
\author[infnferrara]{Marco Romagnoni}
\author[padova,legnaro]{Francesco Sgarbossa}
\author[frascati]{Mattia Soldani}
\author[infnferrara]{Alexei Sytov}
\author[ferrara]{Melissa Tamisari}
\author[minsk]{Victor Tikhomirov}
\author[bicocca]{Erik Vallazza}
\author[padova,legnaro]{Davide Valzani}
\author[como,bicocca]{Giorgio Zuccalà}

\address[infnferrara]{INFN, Sezione di Ferrara, Ferrara, Italy}
\address[como]{Università degli Studi dell'Insubria, Como, Italy}
\address[bicocca]{INFN, Sezione di Milano Bicocca, Milano, Italy}
\address[minsk]{INP, Belarusian State University, Minsk, Belarus}
\address[padova]{Università degli Studi di Padova, Padova, Italy}
\address[legnaro]{INFN, Laboratori Nazionali di Legnaro, Legnaro, Italy}
\address[ferrara]{Università degli Studi di Ferrara, Ferrara, Italy}
\address[infntrieste]{INFN, Sezione di Trieste, Trieste, Italy}
\address[trieste]{Università degli Studi di Trieste, Trieste, Italy}
\address[frascati]{INFN, Laboratori Nazionali di Frascati, Frascati, Italy}

\begin{abstract}
Electromagnetic calorimeters used in high-energy physics and astrophysics rely heavily on high-Z inorganic scintillators, such as lead tungstate (PbWO$_4$ or PWO). The crystalline structure and lattice orientation of inorganic scintillators are frequently underestimated in detector design, even though it is known that the crystalline lattice strongly modifies the features of the electromagnetic processes inside the crystal.
A novel method has been developed for precisely bonding PWO crystals with aligned atomic planes within 100~$\mu$rad, exploiting X-ray diffraction (XRD) to accurately measure miscut angles.
This method demonstrates the possibility to align a layer of  crystals along the same crystallographic direction, opening a new technological path towards the development of next-generation electromagnetic calorimeters.
\end{abstract}


\end{frontmatter}


\section{INTRODUCTION}
\label{sec1}
High-Z inorganic scintillators like lead tungstate (PWO) are widely used in the design of homogeneous electromagnetic calorimeters for high-energy physics (HEP) and astrophysics~\cite{gianotti2003}. Understanding the crystalline nature and lattice orientation is essential as they strongly modify the features of the electromagnetic processes inside the crystal~\cite{Baryshevsky, baier1998electromagnetic}: when a particle moves close to one of the strings (axes) of atoms in the lattice, it experiences an intense electromagnetic field~\cite{uggerhoj2005}, that induces an enhancement of the radiation emission~\cite{soldani2024acceleration, Bandiera23, bandiera2018,soldanithesis, selmithesis} and pair production probability~\cite{soldani2023, pmgthesis} with respect to the Bethe-Heitler description typical of amorphous media~\cite{kimball1985}. The enhancement of radiation and pair production caused by the strong crystalline field results in the acceleration of the electromagnetic shower development~\cite{Bandiera19}, leading to potential advancements in calorimeter resolution, photon detection efficiency, and particle identification capabilities~\cite{pmg_discrimination}. This enhancement extends over an angular range of more than 1 mrad (within which it is maximal), with a detectable effect up to 1\textdegree~\cite{soldani2024acceleration}. The development of oriented detectors are promising for applications in forward geometry in accelerator based experiments and space-borne gamma-ray telescopes~\cite{Bandiera23}. For instance, in case of fixed-target experiments, these effects lead to an improvement of the shower containment in the active volume of the detector, which will result in a major improvement of its energy resolution and a better discrimination of hadronic and electromagnetic signals. In this work, the feasibility of bonding a $3\times 3$ matrix of PWO crystals (Figure~\ref{fig 1}) with the same atomic plane alignment within 100~µrad has been demonstrated.

\section{MATERIAL AND METHODS}
\label{sec2}

Nine samples of PWO Ultra-Fast~\cite{Crytur} were purchased from Crytur. A thorough characterization was carried out using a Panalytical High Resolution X-Ray Diffractometer (HR XRD)~\cite{xray} especially equipped with a laser autocollimator~\cite{GERMOGLI2017308}.
The miscut angle, i.e. the angle between the lattice planes and the sample surface, was measured at the center of each of the crystals with microradian precision and accuracy using the procedure described in~\cite{GERMOGLI2017308, romagnoni2022bent} and previously employed in~\cite{Mazzolari2021}.
A second crucial characterization was the variation of the crystallographic axis orientation at different positions on the sample, that was measured along the two 25mm directions to estimate the mosaicity of the crystal.
Such characterizations were performed on both $25\times25$~$mm^{2}$ faces to identify the upstream surface for each sample in order to minimize the contribution coming from mosaicity and the miscut difference between samples. Indeed, large mosaicity would hinder optimal alignment between samples, while misalignment due to miscut angles may result in gaps in the millimeter range between the faces of adjacent crystals, so it is essential to rotate and adjust these faces to reduce gaps.
Before bonding, all lateral surfaces have been coated with a white, radiation-resistant reflective paint (Eljen EJ-510). This ensures optimal light collection and cancels light loss through the other surfaces, enhancing the overall efficiency of the scintillator. 
During the alignment and bonding procedure, samples were mounted on two separate systems of rotation and linear optomechanic stages via a finely engineered vacuum fixture. 

\begin{figure}
    \centering
    \includegraphics[width=0.9\linewidth]{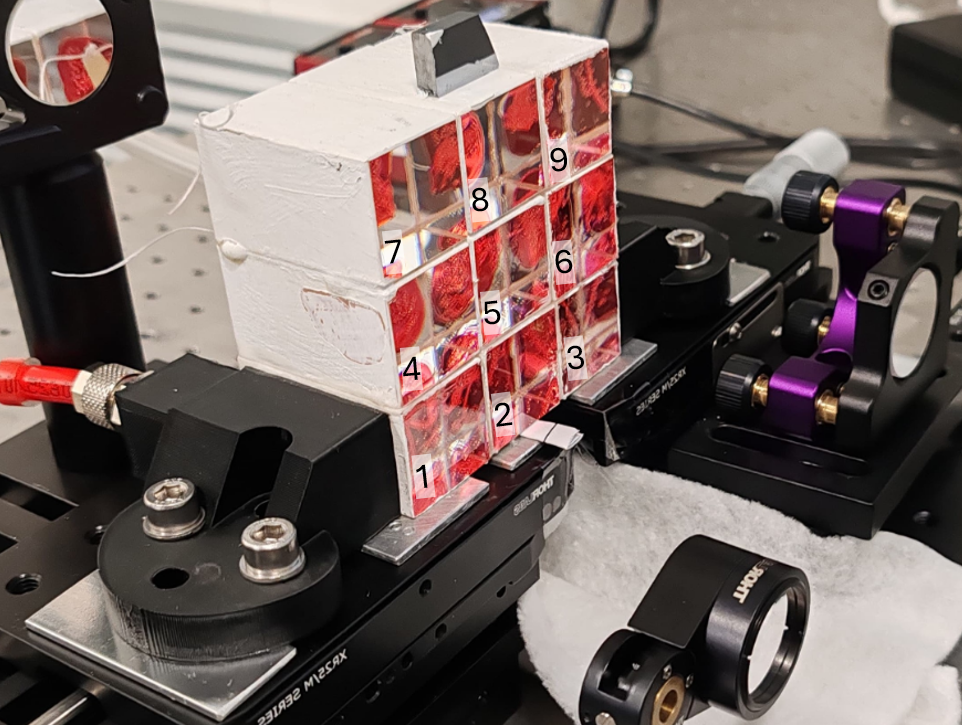}
    \caption{$3\times 3$ matrix formed by oriented crystals with each crystal numbered}
    \label{fig 1}
\end{figure}

A wide-field-of-view laser interferometer (Zygo Verifire HDX~\cite{Zygo}) was employed to measure simultaneously the profile of the samples' $25\times 25\times 45$~$mm^{3}$ surfaces with nanometric precision and accuracy, hence allowing estimation of their reciprocal inclination with few micro-radian uncertainty. Thanks to the previously measured miscut angles, the angle between atomic planes of any pair of crystals can be estimated. Once axes alignment was achieved, an epoxy resin was applied on lateral surfaces and the samples were put in contact. To minimize displacement during curing, resin was applied at the gel point and active adjustment was carried out during initial solidification. Assembly of the nine crystals, with each crystal numbered accordingly, to a $3\times 3$ matrix resulted in the prototype compact calorimeter of Figure~\ref{fig 1}.

\section{RESULTS}
\label{sec3}

\begin{figure}
    \centering
    \includegraphics[width=0.9\linewidth]{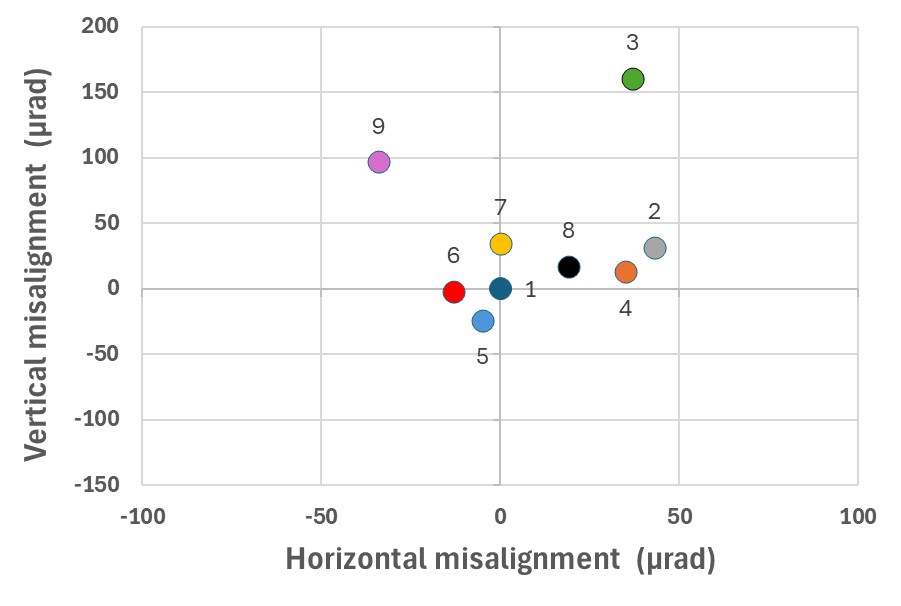}
    \caption{Variation of the horizontal and vertical angle relative to the ideal bonding angle for each crystal. }
    \label{fig 2}
\end{figure}

All PWO crystals for both the $25\times25$~$mm^{2}$ faces present miscut angles within a range of 5000~$\mu$rad, and a crystal mosaicity up to 300~$\mu$rad. Based on these two quantities, the best surface of each crystal has been selected to be used as a reference during the bonding phase. 
The alignment achieved between crystals in the final $3\times 3$  matrix, taking crystal 1 as reference, is shown in Figure~\ref{fig 2}. It is clearly visible that almost all crystals have misalignment lower than 50 ~$\mu$rad, well below the approximately 1 mrad range of the maximum electromagnetic shower acceleration effect observed in PWO. The measurements have been repeated after several months and after handling and transportation, highlighting the robustness of the assembly and absence of any measurable variation in the alignment of the crystals. These results demonstrate that our bonding process has achieved a high level of precision and stability over time.

\section{CONCLUSIONS}
\label{sec4}

In this study, we have developed and tested a novel bonding method for crystals, aimed at obtaining an oriented matrix of PWO. Our method has demonstrated the capability to align a set of crystals with a typical precision and accuracy within a few tens of microradians. Considering that the enhancement of phenomena of interest, such as radiation and pair production, has an angular acceptance on the order of several milliradians, our achieved alignment falls well within the desired region of interest. These results pave the way for the realization of ultra-compact electromagnetic calorimeters based on oriented crystals, which may have significant applications in fixed target experiments and space-borne telescopes~\cite{Bandiera23}.

\section{Aknowledgment}
This work was funded by INFN CSN5 OREO, by the Italian Ministry of University and Research (PRIN 2022Y87K7X) and by the EU Commission (GA 101046458, GA 101032975, GA 101046448). 

  \bibliographystyle{elsarticle-num-names} 
  \bibliography{bibfileTemplate}

\begin{thebibliography}{20}
\expandafter\ifx\csname natexlab\endcsname\relax\def\natexlab#1{#1}\fi
\providecommand{\url}[1]{\texttt{#1}}
\providecommand{\href}[2]{#2}
\providecommand{\path}[1]{#1}
\providecommand{\DOIprefix}{doi:}
\providecommand{\ArXivprefix}{arXiv:}
\providecommand{\URLprefix}{URL: }
\providecommand{\Pubmedprefix}{pmid:}
\providecommand{\doi}[1]{\href{http://dx.doi.org/#1}{\path{#1}}}
\providecommand{\Pubmed}[1]{\href{pmid:#1}{\path{#1}}}
\providecommand{\bibinfo}[2]{#2}
\ifx\xfnm\relax \def\xfnm[#1]{\unskip,\space#1}\fi
\bibitem[{Fabjan and Gianotti(2003)}]{gianotti2003}
\bibinfo{author}{C.~W. Fabjan}, \bibinfo{author}{F.~Gianotti},
\newblock \bibinfo{title}{Calorimetry for particle physics},
\newblock \bibinfo{journal}{Rev. Mod. Phys.} \bibinfo{volume}{75} (\bibinfo{year}{2003}) \bibinfo{pages}{1243--1286}. \DOIprefix\doi{10.1103/RevModPhys.75.1243}.
\bibitem[{Baryshevsky et~al.(2017)}]{Baryshevsky}
\bibinfo{author}{V.~Baryshevsky}, et~al.,
\newblock \bibinfo{title}{{On the influence of crystal structure on the electromagnetic shower development in the lead tungstate crystals}},
\newblock \bibinfo{journal}{Nucl. Instrum. Methods Phys.} \bibinfo{volume}{402} (\bibinfo{year}{2017}) \bibinfo{pages}{35--39}. \DOIprefix\doi{doi.org/10.1016/j.nimb.2017.02.066}.
\bibitem[{Baier et~al.(1998)Baier, Katkov, and Strakhovenko}]{baier1998electromagnetic}
\bibinfo{author}{V.~N. Baier}, \bibinfo{author}{V.~M. Katkov}, \bibinfo{author}{V.~V. Strakhovenko}, \bibinfo{title}{Electromagnetic processes at high energies in oriented single crystals}, \bibinfo{publisher}{World Scientific}, \bibinfo{year}{1998}. \DOIprefix\doi{10.1142/2216}.
\bibitem[{Uggerh\o{}j(2005)}]{uggerhoj2005}
\bibinfo{author}{U.~I. Uggerh\o{}j},
\newblock \bibinfo{title}{{The interaction of relativistic particles with strong crystalline fields}},
\newblock \bibinfo{journal}{Rev. Mod. Phys.} \bibinfo{volume}{77} (\bibinfo{year}{2005}) \bibinfo{pages}{1131--1171}. \DOIprefix\doi{10.1103/RevModPhys.77.1131}.
\bibitem[{Soldani et~al.(2024)}]{soldani2024acceleration}
\bibinfo{author}{M.~Soldani}, et~al.,
\newblock \bibinfo{title}{Acceleration of electromagnetic shower development and enhancement of light yield in oriented scintillating crystals},
\newblock \bibinfo{journal}{arXiv preprint arXiv:2404.12016}  (\bibinfo{year}{2024}). \URLprefix \url{https://arxiv.org/abs/2404.12016}.
\bibitem[{Bandiera et~al.(2023)}]{Bandiera23}
\bibinfo{author}{L.~Bandiera}, et~al.,
\newblock \bibinfo{title}{A highly-compact and ultra-fast homogeneous electromagnetic calorimeter based on oriented lead tungstate crystals},
\newblock \bibinfo{journal}{Front. Phys.} \bibinfo{volume}{11} (\bibinfo{year}{2023}) \bibinfo{pages}{10.3389/fphy.2023.1254020}. \URLprefix \url{https://www.frontiersin.org/articles/10.3389/fphy.2023.1254020}. \DOIprefix\doi{10.3389/fphy.2023.1254020}.
\bibitem[{Bandiera et~al.(2018)}]{bandiera2018}
\bibinfo{author}{L.~Bandiera}, et~al.,
\newblock \bibinfo{title}{Strong reduction of the effective radiation length in an axially oriented scintillator crystal},
\newblock \bibinfo{journal}{Phys. Rev. Lett.} \bibinfo{volume}{121} (\bibinfo{year}{2018}) \bibinfo{pages}{021603}. \DOIprefix\doi{10.1103/PhysRevLett.121.021603}.
\bibitem[{Soldani(2023)}]{soldanithesis}
\bibinfo{author}{M.~Soldani}, \bibinfo{title}{{Innovative applications of strong crystalline field effects to particle accelerators and detectors}}, \bibinfo{type}{Master's thesis}, Università degli Studi di Ferrara, \bibinfo{year}{2023}. \URLprefix \url{https://cds.cern.ch/record/2864634}.
\bibitem[{Selmi(2022)}]{selmithesis}
\bibinfo{author}{A.~Selmi}, \bibinfo{title}{{Electromagnetic Shower Development in Oriented Crystals}}, \bibinfo{type}{Master's thesis}, Università degli Studi dell'Insubria, \bibinfo{year}{2022}. \URLprefix \url{https://cds.cern.ch/record/2814582}.
\bibitem[{Soldani et~al.(2023)}]{soldani2023}
\bibinfo{author}{M.~Soldani}, et~al.,
\newblock \bibinfo{title}{{Strong enhancement of electromagnetic shower development induced by high-energy photons in a thick oriented tungsten crystal}},
\newblock \bibinfo{journal}{Eur. Phys. J. C} \bibinfo{volume}{83} (\bibinfo{year}{2023}) \bibinfo{pages}{101}. \DOIprefix\doi{10.1140/epjc/s10052-023-11247-x}.
\bibitem[{Monti-Guarnieri(2023)}]{pmgthesis}
\bibinfo{author}{P.~Monti-Guarnieri}, \bibinfo{title}{{Beamtest characterization of oriented crystals for the KLEVER Small Angle Calorimeter}}, \bibinfo{type}{Master's thesis}, Università degli Studi dell'Insubria, \bibinfo{year}{2023}. \URLprefix \url{https://cds.cern.ch/record/2850897}.
\bibitem[{Kimball and Cue(1985)}]{kimball1985}
\bibinfo{author}{J.~C. Kimball}, \bibinfo{author}{N.~Cue},
\newblock \bibinfo{title}{Quantum electrodynamics and channeling in crystals},
\newblock \bibinfo{journal}{Phys. Rep.} \bibinfo{volume}{125} (\bibinfo{year}{1985}) \bibinfo{pages}{69--101}. \DOIprefix\doi{10.1016/0370-1573(85)90021-3}.
\bibitem[{Bandiera et~al.(0189)}]{Bandiera19}
\bibinfo{author}{L.~Bandiera}, et~al.,
\newblock \bibinfo{title}{Compact electromagnetic calorimeters based on oriented scintillator crystals},
\newblock \bibinfo{journal}{Nuclear Inst. and Methods in Physics Research, A} \bibinfo{volume}{936} (\bibinfo{year}{20189}) \bibinfo{pages}{124--126}. \DOIprefix\doi{doi.org/10.1016/j.nima.2018.07.085}.
\bibitem[{Monti-Guarnieri(2024)}]{pmg_discrimination}
\bibinfo{author}{P.~Monti-Guarnieri},
\newblock \bibinfo{title}{{Particle identification with a longitudinally segmented homogeneous calorimeter composed of oriented crystals}}  (\bibinfo{year}{2024}). \DOIprefix\doi{doi.org/10.48550/arXiv.2405.11302}, \bibinfo{note}{{Submitted to arXiv}}.
\bibitem[{Crytur(2024)}]{Crytur}
\bibinfo{author}{Crytur}, \bibinfo{title}{integrated crystal based solution}, \bibinfo{year}{2024}. \bibinfo{note}{Available at \url{https://www.crytur.com/}}.
\bibitem[{Panalytical(2024)}]{xray}
\bibinfo{author}{Panalytical}, \bibinfo{title}{X-ray Diffractometer}, \bibinfo{year}{2024}. \bibinfo{note}{Available at \url{https://www.malvernpanalytical.com/en/products/technology/xray-analysis/x-ray-diffraction}}.
\bibitem[{Germogli et~al.(2017)}]{GERMOGLI2017308}
\bibinfo{author}{G.~Germogli}, et~al.,
\newblock \bibinfo{title}{Bent silicon strip crystals for high-energy charged particle beam collimation},
\newblock \bibinfo{journal}{Nuclear Instruments and Methods in Physics Research Section B: Beam Interactions with Materials and Atoms} \bibinfo{volume}{402} (\bibinfo{year}{2017}) \bibinfo{pages}{308--312}. \DOIprefix\doi{https://doi.org/10.1016/j.nimb.2017.03.053}, \bibinfo{note}{proceedings of the 7th International Conference Channeling 2016: Charged $\&$ Neutral Particles Channeling Phenomena}.
\bibitem[{Romagnoni et~al.(2022)}]{romagnoni2022bent}
\bibinfo{author}{M.~Romagnoni}, et~al.,
\newblock \bibinfo{title}{Bent crystal design and characterization for high-energy physics experiments},
\newblock \bibinfo{journal}{Crystals} \bibinfo{volume}{12} (\bibinfo{year}{2022}) \bibinfo{pages}{1263}. \DOIprefix\doi{https://doi.org/10.3390/cryst12091263}.
\bibitem[{Mazzolari et~al.(2021)}]{Mazzolari2021}
\bibinfo{author}{A.~Mazzolari}, et~al.,
\newblock \bibinfo{title}{{Silicon crystals for steering high-intensity particle beams at ultrahigh-energy accelerators}},
\newblock \bibinfo{journal}{Phys. Rev. Res} \bibinfo{volume}{3} (\bibinfo{year}{2021}) \bibinfo{pages}{013108}. \DOIprefix\doi{10.1103/PhysRevResearch.3.013108}.
\bibitem[{Zygo(2024)}]{Zygo}
\bibinfo{author}{Zygo}, \bibinfo{title}{interferometers}, \bibinfo{year}{2024}. \bibinfo{note}{Available at \url{https://www.zygo.com/products/metrology-systems/laser-interferometers}}.

\end{thebibliography}

\end{document}